\documentclass[showpacs,aps,prd,nofootinbib,floatfix,amsmath,amssymb,twocolumn]{revtex4}
\usepackage{graphicx}
\usepackage{hyperref}
\usepackage{float}
\restylefloat{table}
\begin{document}

\makeatletter
\newbox\slashbox \setbox\slashbox=\hbox{$/$}
\newbox\Slashbox \setbox\Slashbox=\hbox{\large$/$}
\def\pFMslash#1{\setbox\@tempboxa=\hbox{$#1$}
  \@tempdima=0.5\wd\slashbox \advance\@tempdima 0.5\wd\@tempboxa
  \copy\slashbox \kern-\@tempdima \box\@tempboxa}
\def\pFMSlash#1{\setbox\@tempboxa=\hbox{$#1$}
  \@tempdima=0.5\wd\Slashbox \advance\@tempdima 0.5\wd\@tempboxa
  \copy\Slashbox \kern-\@tempdima \box\@tempboxa}
\def\FMslash{\protect\pFMslash}
\def\FMSlash{\protect\pFMSlash}
\def\miss#1{\ifmmode{/\mkern-11mu #1}\else{${/\mkern-11mu #1}$}\fi}
\makeatother

\title{Implications of a nonvanishing $Z \gamma \gamma$ vertex on the $H\to \gamma \gamma \gamma$ decay}
\author{A. Cordero--Cid$^{(a)}$, M. A. L\' opez-Osorio$^{(b)}$, E. Mart\'\i nez-Pascual$^{(b)}$, and J. J. Toscano$^{(b)}$}

\affiliation{
$^{(a)}$ Facultad de Ciencias de la Electr\' onica, Benem\' erita Universidad
Aut\' onoma de Puebla, Blvd. 18 Sur y Av. San Claudio, 72590, Puebla, Pue., M\' exico.\\
$^{(b)}$Facultad de Ciencias F\'{\i}sico Matem\'aticas,
Benem\'erita Universidad Aut\'onoma de Puebla, Apartado Postal
1152, Puebla, Puebla, M\'exico.}

\begin{abstract}
The $Z\to \gamma \gamma$ and $H\to \gamma \gamma \gamma$ decays are strictly forbidden in the Standard Model, but they can be induced by theories that violate Lorentz symmetry or the CPT theorem. By assuming that a nonvanishing $Z\gamma \gamma$ vertex is induced in some context of new physics, and by analyzing the reaction $H\to \gamma Z^*\to \gamma \gamma \gamma$ in the $Z$ resonance, we obtain an estimation for the branching ratio of the $H\to \gamma \gamma \gamma$ decay. Specifically, it is found that $BR(H\to \gamma \gamma \gamma)\lesssim BR(H\to \gamma Z)BR_{Exp}(Z\to \gamma \gamma)$, which is of the  order of $10^{-7}$.
\end{abstract}


\maketitle
Processes that are forbidden or highly suppressed constitute a natural framework to test any new physics lying beyond the standard model (SM). Within  this category, one finds the decays $Z\to \gamma \gamma$ and $H\to \gamma \gamma \gamma$.

The  decay of the $Z$ gauge boson into a pair of photons or gluons is forbidden in standard field theories due to angular momentum conservation and Bose statistics. This fact, which is known as the Landau-Yang theorem~\cite{LYT}, applies to any transition of a massive vector boson into two massless vector particles\footnote{The requirement of two massless vector particles in the final state is crucial in the Landau-Yang theorem, as decays of the form $Z'\to Z\gamma$ can occur in theories that predict a new $Z'$ gauge boson~\cite{ZpZg}.}. However, this type of decays can occur in some SM extensions that predict Lorentz violation at very small distances or very high energies, such as noncommutative field theories~\cite{NCSM} or the Standard Model Extension (SME)~\cite{SME}. This latter is an effective theory that incorporates Lorentz violation and CPT nonconservation in a model-independent manner; it was motivated from specific scenarios in the context of string theory~\cite{ST} or general relativity with spontaneous symmetry breaking~\cite{GR}, but the SME is beyond these specific ideas due to its generality, which is the main advantage of effective field theories. About the noncommutative Standard Model (NCSM), it was shown that it is a subset of the SME~\cite{NCSME}. It is in the specific context of the NCSM, that the $Z$ boson decay into pairs of photons or gluons has already been studied~\cite{ZggNCSM}. On the other hand, the  $H\to \gamma \gamma \gamma$ decay is forbidden due to charge conjugation conservation. So, the prohibition of the $Z\to \gamma \gamma$ and $H\to \gamma \gamma \gamma$ decays is linked to fundamental principles of quantum field theory.

The recent discovery of the Higgs boson at the Large Hadron Collider (LHC)~\cite{ATLAS,CMS} constitutes a strong incentive to investigate forbidden Higgs boson decays, as the aforementioned $H\to \gamma \gamma \gamma$ decay, which, to our best knowledge, has not been considered in the literature. Forbidden processes constitute sharp windows through which new physics effects may show up. Unlike those processes that fall into the category of rare or very suppressed decays, the forbidden ones are intrinsically relevant to test fundamental concepts. This is the case of the  $Z\to \gamma \gamma$ and $H\to \gamma \gamma \gamma$ decays, which, necessarily arise as a consequence of Lorentz violation and charge conjugation nonconservation, respectively. In a wider scenario, both types of decays may arise from the violation of the fundamental CPT theorem, which in turn implies Lorentz violation~\cite{CPTVvsLV}. The interesting possibility that these decays can be induced by the same source of new physics cannot be discarded. In this brief report, we show that this connection is possible if a nonvanishing $Z\gamma \gamma$ vertex is induced by some source of Lorentz violation. We also obtain an estimation for the branching ratio $BR(H\to \gamma \gamma \gamma)$ using the experimental constraint on  $BR(Z\to \gamma \gamma)$.

Although in this work we will not refer to a specific scenario of CPT or Lorentz violation (NCSM or SME) to predict the branching ratio for the $H\to \gamma \gamma \gamma$ decay, except to make use of the assumption that a nonvanishing $Z\gamma \gamma$ vertex exits, for the sake of clarity, it is convenient to present a brief discussion on two possible scenarios in which the  $Z\to \gamma \gamma$ and $H\to \gamma \gamma \gamma$ decays could arise. In one of them, the source that induces these decays is CPT violation, where Lorentz violation is implicit~\cite{CPTVvsLV}. The other scenario considers that the CPT theorem is preserved, but the Lorentz symmetry is violated.

\textbf{The CPT-odd scenario.} A typical CPT-violating scenario~\cite{CPTV}, which has been the subject of important attention in the literature~\cite{CJ,PV}, is characterized by an interaction of the form
\begin{equation}
{\cal L}_{CPT-odd}=b_\mu \bar{\psi}\gamma_5 \gamma^\mu \psi\, ,
\end{equation}
where $\psi$ stands for the spinor field characterizing a lepton or quark and $b_\mu$ is assumed to be real\footnote{This is to consider that it may arise from spontaneous symmetry breaking of the Lorentz symmetry at very high energies and also to ensure that the underlying theory is hermitian.}. In this scenario, the $Z\to \gamma \gamma$ and $H\to \gamma \gamma \gamma$ decays can be induced by triangle and box diagrams, respectively, in which  charged leptons and quarks circulate, and whose propagators incorporate insertions of the CPT-violating $ib_\mu \gamma_5 \gamma^\mu$ vertex. Of course, the amplitudes for these processes are exactly zero for $b=0$, as predicted by the SM, but nonvanishing contributions proportional to $b$ can arise. At first order in $b$, a possible gauge and Lorentz structure of the $Z\gamma \gamma$ vertex would be given by the following dimension-five operator
\begin{equation}
{\cal O}_{Z\gamma \gamma}=\frac{F(m^2_f,m_{Z}^{2})}{m^2_f}\, b^\alpha Z_\alpha F_{\mu \nu}F^{\mu \nu}\, ,
\end{equation}
where $F(m^2_f,m^2_Z)$ is a dimensionless loop function, $m_f$ is the mass of the virtual fermion, and $F_{\mu \nu}$ is the electromagnetic curvature. This $U_e(1)$-invariant operator is in turn induced by a linear combination of the dimension-seven $SU_C(3)\times SU_L(2)\times U_Y(1)$-invariant  operators $b^\alpha (\Phi^\dag D_\alpha \Phi-\textrm{h.c.})\textrm{Tr}(W_{\mu \nu}W^{\mu \nu})$ and  $b^\alpha (\Phi^\dag D_\alpha \Phi-\textrm{h.c.})B_{\mu \nu}B^{\mu \nu}$, where $\Phi$ is the Higgs doublet, whereas $W_{\mu \nu}$ and $B_{\mu \nu}$ are the respective $SU_L(2)$ and $U_Y(1)$ curvatures. As far as the $H\gamma \gamma \gamma$ vertex is concerned, a representative operator characterizing this interaction would be
\begin{equation}
{\cal O}_{H\gamma \gamma \gamma}=\frac{A(m^2_f,m^2_H)}{m^5_f}\, b^\alpha \partial^\beta H F_{\alpha \beta}\tilde{F}_{\mu \nu}F^{\mu \nu}\, ,
\end{equation}
where $\tilde{F}_{\mu \nu}=(1/2)\epsilon_{\mu \nu \alpha \beta}F^{\alpha \beta}$ is the electromagnetic dual tensor and $ A(m_{f}^{2},m_{H}^{2}) $ is a dimensionless loop function. Although this dimension-eight operator is manifestly invariant only under the $U_e(1)$ group, it arises indeed as a linear combination from the dimension-nine  $SU_C(3)\times SU_L(2)\times U_Y(1)$-invariant operators $b^\alpha (\Phi^\dag D^\beta \Phi+\textrm{h.c.})\textrm{Tr}(W_{\alpha \beta}\tilde{W}_{\mu \nu}W^{\mu \nu})$, $b^\alpha (\Phi^\dag D^\beta \Phi+\textrm{h.c.})\textrm{Tr}(\tilde{W}_{\mu \nu}W^{\mu \nu})B_{\alpha \beta}$, and $b^\alpha (\Phi^\dag D^\beta \Phi+\textrm{h.c.})\tilde{B}_{\mu \nu}B^{\mu \nu}B_{\alpha \beta}$. Of course, there must be several independent $SU_C(3)\times SU_L(2)\times U_Y(1)$-invariant operators characterizing both the $Z\gamma \gamma$ and $H\gamma \gamma \gamma$ vertices (including those that may be generated at higher orders in $b$), but the ones displayed above are enough for illustrative purposes.

\textbf{The Lorentz-violating CPT-even scenario.} One scenario that presents Lorentz violation with CPT conservation is the NCSM~\cite{NCSM}. In this model, Lorentz violation arises via a totally antisymmetric real object $\theta^{\mu \nu}$, which characterizes the noncommutativity of the space-time, $[x^\mu, \, x^\nu]=i\theta^{\mu \nu}$. One feature of the NCSM is the presence of the $Z\gamma \gamma$ and $\gamma \gamma \gamma$ vertices at first order in $\theta^{\mu \nu}$~\cite{ZggNCSM}. However, as it was shown in~\cite{NCSME}, the NCSM arises naturally as a subset of the SME~\cite{SME}. In this more general description of Lorentz violation given by the SME, vertices $Z\gamma \gamma$ and $\gamma \gamma \gamma$ emerge from the following dimension-six $SU_C(3)\times SU_L(2)\times U_Y(1)$-invariant operators
\begin{eqnarray}
{\cal O}^{WB(1)}_{CPT-even}&=&\frac{\alpha_{WB(1)}}{\Lambda^2}\, b^{\alpha \beta}\textrm{Tr}(W_{\alpha \beta}W^{\mu \nu})B_{\mu \nu}\, ,\\
{\cal O}^{WB(2)}_{CPT-even}&=&\frac{\alpha_{WB(2)}}{\Lambda^2}\, b^{\alpha \beta}\textrm{Tr}(W_{\mu \nu}W^{\mu \nu})B_{\alpha \beta}\, ,\\
{\cal O}^{WB(3)}_{CPT-even}&=&\frac{\alpha_{WB(3)}}{\Lambda^2}\, b^{\alpha \beta}\textrm{Tr}(W_{\alpha \mu}W^{\nu \mu})B_{\nu \beta}\, ,
\end{eqnarray}
where the $\alpha_{WB(i)}$ dimensionless coefficients depend on the details of the underlying theory and $\Lambda$ is the scale characterizing the new physics effects. The presence of the dimensionless $b_{\alpha \beta}$ object may be a consequence of noncommutativity of the space-time ($b_{\alpha \beta}=\theta_{\alpha \beta}/\Lambda^2$), or the vacuum expectation value of a tensor field $B_{\alpha \beta}(x)$ responsible for a spontaneous symmetry breaking of the Lorentz group $SO(1,3)$, or other unknown sources.

In this context, the $H\gamma \gamma \gamma$ vertex arises from the dimension-eight gauge invariant operators
\begin{align}
{\cal O}^{\Phi WB(1)}_{CPT-even} &=\frac{\alpha_{\Phi WB(1)}}{\Lambda^4}\, b^{\alpha \beta} (\Phi^\dag \Phi)\textrm{Tr}(W_{\alpha \beta}W^{\mu \nu})B_{\mu \nu}\, ,\\
{\cal O}^{\Phi WB(2)}_{CPT-even}&=\frac{\alpha_{\Phi WB(2)}}{\Lambda^4}\, b^{\alpha \beta}(\Phi^\dag \Phi) \textrm{Tr}(W_{\mu \nu}W^{\mu \nu})B_{\alpha \beta}\, ,\\
{\cal O}^{\Phi WB(3)}_{CPT-even}&=\frac{\alpha_{\Phi WB(3)}}{\Lambda^4}\, b^{\alpha \beta}(\Phi^\dag \Phi) \textrm{Tr}(W_{\alpha \mu}W^{\nu \mu})B_{\nu \beta}\, .
\end{align}
It should be noticed that the $H\gamma \gamma \gamma$ vertex is naturally suppressed by a factor of $(v/\Lambda)^2$ with respect to the $Z\gamma \gamma$ one, where $v$ is the Fermi scale.

\begin{figure}[htbp!]
\centering
\includegraphics[scale=.4]{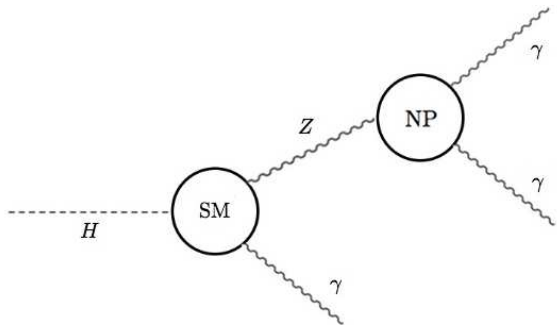}
\caption{Typical diagram characterizing the $H\to \gamma Z^*\to \gamma \gamma \gamma$ decay. In this figure, SM and NP stand for Standard Model contribution and New Physics effects, respectively. }
\label{H3g}
\end{figure}

Having discussed some possible sources of new physics effects responsible for the presence of the $Z\gamma \gamma$ and $H\gamma \gamma \gamma$ vertices, we proceed to obtain an estimation for the branching ratio of the $H\to \gamma \gamma \gamma$ decay. We will derive this with absolute independence of the details of a specific model, only the existence of a nonvanishing $Z\gamma \gamma$ vertex will be assumed. In fact, if a nonvanishing $Z\gamma \gamma$ vertex is generated in some theory beyond the SM, the decay $H\to \gamma \gamma \gamma$ is naturally induced via the one-loop SM $H\gamma Z^*$ vertex (with $Z^*$ denoting a virtual $Z$ gauge boson) via the reaction
\begin{equation}
H\to \gamma Z^*\to \gamma \gamma \gamma\, ,
\end{equation}
shown in Fig. \ref{H3g}. We will derive a result for the branching ratio associated with this decay in the narrow width approximation, which is valid as long as $\Gamma_Z\ll m_Z$. This assumption is crucial for our subsequent discussion. The squared of the invariant amplitude that arises from these types of diagrams can be written as
\begin{equation}
|{\cal M}(H\to \gamma \gamma \gamma)|^2= D^2|{\cal M}^{H\gamma Z}_\mu (-g^{\mu \nu}+\frac{k^\mu k^\nu}{m^2_Z}){\cal M}^{Z\gamma \gamma}_\nu|^2\, ,
\end{equation}
where
\begin{equation}
D^2=\frac{1}{(k^2-m^2_Z)^2 +m^2_Z\Gamma^2_Z}
\end{equation}
is the squared denominator of the $Z$ propagator. In addition,
\begin{eqnarray}
{\cal M}^{H\gamma Z}_\mu&=&\Gamma^{SM}_{\mu \mu_3}\epsilon^{*\mu_3}(k_3,\lambda_3) \, ,\\
{\cal M}^{Z\gamma \gamma}_\nu&=&\Gamma^{NP}_{\nu \mu_1\mu_2}\epsilon^{*\mu_1}(k_1,\lambda_1)\epsilon^{*\mu_2}(k_2,\lambda_2)\, ,
\end{eqnarray}
where the tensors $\Gamma^{SM}_{\mu \mu_3}(k,k_3)$ and $\Gamma^{NP}_{\nu \mu_1\mu_2}(k,k_1,k_2)$ represent the vertices of the $H\gamma Z^*$ and $Z^*\gamma \gamma$ couplings (see Fig. \ref{H3g}), whose specific form is not needed here. In these expressions, the quantities $\epsilon^{*\mu_i}(k_i,\lambda_i)$ are the polarization vectors of the final photons.

The standard narrow width approximation\footnote{This approximation is associated with the case
 in which the two subprocesses in consideration are linked by a spin $0$ particle.} is obtained when a polarization correlation between the $H\to \gamma Z$ and $Z\to \gamma \gamma$ subprocesses is neglected. This is a good approximation indeed, as it has been shown recently~\cite{NWA} in a more general context, the fact that polarization correlation effects can be disregarded in this type of processes. Under this assumption, the three-body phase space can be written as the product of the phase spaces associated to the $H\to \gamma Z$ and $Z\to \gamma \gamma$ subprocesses as follows:
\begin{equation}
d\Phi_3(H\to \gamma \gamma \gamma)=d\Phi_2(H\to \gamma Z)\frac{dk^2}{2\pi}d\Phi_2(Z\to \gamma \gamma)\,
\end{equation}
and
\begin{equation}
\lim_{\Gamma\to 0}D^2=\frac{\pi}{m_Z\Gamma_Z}\delta(k^2-m^2_Z)\, .
\end{equation}
\begin{widetext}
On the other hand, the numerator of the $Z$ propagator can be replaced by its polarization vector, to obtain
\begin{align}
|{\cal M}^{H\gamma Z}_\mu (-g^{\mu \nu}+\frac{k^\mu k^\nu}{m^2_Z}){\cal M}^{Z\gamma \gamma}_\nu|^2 & = |{\cal M}^{H\gamma Z}_\mu \sum_{\lambda}\epsilon^{*\mu}(k,\lambda)\epsilon^\nu(k,\lambda){\cal M}^{Z\gamma \gamma}_\nu|^2\,\nonumber \\
& \to |{\cal M}(H\to \gamma Z)|^2\frac{1}{3}|{\cal M}(Z\to \gamma \gamma)|^2\, .
\end{align}
\end{widetext}
Taking into account that there are three diagrams of the type of the one display in Fig. \ref{H3g}, the branching ratio for the $H\to \gamma \gamma \gamma$ decay can be written as
\begin{equation}
BR(H\to \gamma \gamma \gamma)=BR(H\to \gamma Z)BR(Z\to \gamma \gamma)\, .
\end{equation}
 Using the SM prediction for the branching ratio of the $H\to \gamma Z$ decay, given by $BR(H\to \gamma Z)=1.42\times 10^{-3}$, we obtain the following bound for the branching ratio of the Higgs boson decay into three photons
\begin{equation}
BR(H\to \gamma \gamma \gamma)<(1.42\times 10^{-3})BR_{Exp}(Z\to \gamma \gamma)\, ,
\end{equation}
being $BR_{Exp}(Z\to \gamma \gamma)$ the experimental limit on the branching ratio of the $Z\to \gamma \gamma$ decay. The current experimental bound for this decay is~\cite{PDG} $BR_{Exp}(Z\to \gamma \gamma)<5.2\times 10^{-5}$, which results in the following estimation for $BR(H\to \gamma \gamma \gamma)$:
\begin{equation}
BR(H\to \gamma \gamma \gamma)<7.38\times 10^{-8}\, .
\end{equation}
It is interesting to compare the branching ratio for the $H\to \gamma \gamma \gamma$ decay calculated in this way, with the one that could be obtained in more specific scenarios. In the CPT-odd scenario discussed above, the $\gamma \gamma \gamma$ vertex is not generated at one-loop level~\cite{QED}, so the $H\to \gamma \gamma \gamma$ decay can be induced at one-loop level via box diagrams and also via the one-loop vertices $H\gamma Z$ and $Z\gamma \gamma$ connected by the $Z$ gauge boson, whose resonant effect has been considered above in a model-independent way. Radiative corrections tell us that the box contribution can be depreciated against the $Z$ resonant effect. So, one expects that in general, if both vertices, $H\gamma \gamma \gamma$ and $Z\gamma \gamma$, are induced at one-loop level, the best estimation for the $BR(H\to \gamma \gamma \gamma)$ would be obtained from the reducible process $H\to \gamma Z^*\to \gamma \gamma \gamma$ with the $Z$ gauge boson in resonance. On the other hand, in a scenario where both the $Z\gamma \gamma$ and $\gamma \gamma \gamma$ vertices are induced, as it is the case of the NCSM or the SME, the reducible process $H\to \gamma \gamma^*\to \gamma \gamma \gamma$ may be comparable to the $H\to \gamma Z^*\to \gamma \gamma \gamma$ one. However, the specific details of the model are needed to perform the calculations. In particular, the corresponding amplitude would depend on specific spatial directions characterized by the constant objets carrying Lorentz indices that characterize this type of theories. In much more specific scenarios, the estimations for the $BR(H\to \gamma \gamma \gamma)$ would depend excessively on the details of the model.

In conclusion, in this paper we have derived a model-independent estimation for the branching ratio of the $H\to \gamma \gamma \gamma$ decay, which was  given in terms of the SM prediction for the $BR(H\to \gamma Z)$ and the experimental limit on the $BR(Z\to \gamma \gamma)$.

\acknowledgments{We acknowledge financial support from CONACYT and A.C.C., E.M.P. and J.J.T. also acknowledge
SNI (M\' exico).}

\end{document}